\documentclass[manuscript]{aastex61}
\usepackage[T1]{fontenc}

\usepackage{bm}        
\usepackage{amssymb, amsmath}   
\usepackage{cancel}
\usepackage{color}

\DeclareMathAlphabet{\mathsfit}{\encodingdefault}{\sfdefault}{m}{sl}
\SetMathAlphabet{\mathsfit}{bold}{\encodingdefault}{\sfdefault}{bx}{sl}
\newcommand{\vect}[1]{\bm{#1}}

\DeclareMathOperator{\sech}{sech}
\newcommand{\tensorsym}[1]{\bm{\mathsfit{#1}}}

\received{\today}
\revised{}
\accepted{}
\submitjournal{ApJ}

\shorttitle{Compression and Shear in Reconnection}
\shortauthors{Li et al.}

\begin{document}

\title{The roles of fluid compression and shear in electron energization during
magnetic reconnection}

\correspondingauthor{Xiaocan Li}
\email{xiaocanli@lanl.gov}

\author[0000-0001-5278-8029]{Xiaocan Li}
\affil{Los Alamos National Laboratory, Los Alamos, NM 87544, USA}

\author{Fan Guo}
\affiliation{Los Alamos National Laboratory, Los Alamos, NM 87544, USA}
\affiliation{New Mexico Consortium, 4200 West Jemez Road, Los Alamos, NM 87544, USA}

\author{Hui Li}
\affiliation{Los Alamos National Laboratory, Los Alamos, NM 87544, USA}

\author{Joachim Birn}
\affiliation{Space Science Institute, Boulder, CO 80301, USA}
\affiliation{Guest Scientist, Los Alamos National Laboratory, Los Alamos, NM 87544, USA}

\begin{abstract}
  Particle acceleration in space and astrophysical reconnection sites is an
  important unsolved problem in studies of magnetic reconnection. Earlier kinetic
  simulations have identified several acceleration mechanisms that are associated
  with particle drift motions. Here, we show that, for sufficiently large systems,
  the energization processes due to particle drift motions can be described as
  fluid compression and shear, and that the shear energization is proportional to
  the pressure anisotropy of energetic particles. By analyzing results from fully
  kinetic simulations, we show that the compression energization dominates the
  acceleration of high-energy particles in reconnection with a weak guide field,
  and the compression and shear effects are comparable when the guide field is
  50\% of the reconnecting component. Spatial distributions of those energization
  effects reveal that reconnection exhausts, contracting islands, and island-merging
  regions are the three most important regions for compression and shear acceleration.
  This study connects particle energization by particle guiding-center drift
  motions with that due to background fluid
  motions, as in the energetic particle transport theory. It provides
  foundations for building particle transport models for large-scale reconnection
  acceleration such as those in solar flares.
\end{abstract}


\section{Introduction} \label{sec:intro}
Magnetic reconnection is a major mechanism that drives the release of magnetic
energy in space and astrophysical plasmas~\citep{Zweibel2009Magnetic}.
For example, magnetic reconnection converts $10\%-50\%$ of the magnetic energy
into plasma kinetic energy within $10^{2-3}$ s~\citep{Lin1976Nonthermal} and heats
solar coronal plasma from $\sim 1$ MK to up to over 30 MK during solar
flares~\citep{Caspi2010RHESSI, Longcope2010Model}. Besides heating, observations
indicate that magnetic reconnection can accelerate about 10\% of
electrons~\citep{Oka2013Kappa, Oka2015Electron} or even the entire electron
population in a solar flare region (more than $10^{36}$ electrons) into a
nonthermal distribution~\citep{Krucker2010Measure, Krucker2014Particle}. Such
efficient particle acceleration over a large-scale reconnection region is an
important unsolved problem in the study of reconnection.

Previous reconnection studies have identified that particles are accelerated
close to the reconnection $X$-point~\citep{Hoshino2001Suprathermal,
Drake2005Production, Fu2006Process, Oka2010Electron, Egedal2012Large,
Egedal2015Double,Wang2016Mechanisms}, in contracting magnetic islands~\citep{Drake2006Electron,
Oka2010Electron}, and also in island-merging regions~\citep{Oka2010Electron,
Liu2011Particle, Drake2013Power, Nalewajko2015Distribution}. At the $X$-points,
particles get accelerated by streaming along the nonideal electric field. In the
contracting and merging magnetic islands, the acceleration closely resembles
to \textit{Fermi}-type processes. In order to clarify the acceleration mechanism,
some recent works calculated the energy gain by summing over particle motions under
the guiding-center approximation and identified curvature drift as the primary
particle acceleration mechanism during reconnection~\citep{Dahlin2014Mechanisms,
Guo2014Formation, Guo2015Particle,Li2015Nonthermally,Zhou2015Electron,Beresnyak2016First}. This
drift acceleration is similar to the \textit{Fermi} process because particles gain energy
proportional to their kinetic energy when they are bouncing between two sides of
a magnetic island and in island-merging regions.

To understand particle acceleration in a large-scale reconnection layer,
an important task is to develop a statistical transport theory that 
includes the primary acceleration mechanisms.
In shock acceleration theory, the Parker transport equation has provided the basic
description for the acceleration and transport of energetic particles in the
shock region, where adiabatic compression is the leading acceleration
mechanism~\citep{Parker1965Passage,Blandford1987Particle}. Acceleration due to
velocity shear and fluid inertia have been considered as higher-order
effects~\citep[e.g.,][]{Earl1988Cosmic, Zank2014Transport}.
Several reconnection studies have attempted
to develop similar kinetic equations to evolve electron distribution. The most
common approach is to derive a reduced kinetic equation from the guiding-center
drift kinetic equation by assuming double-adiabatic invariants:
the magnetic moment and the parallel action integral~\citep{Drake2006Electron,
Drake2013Power, Egedal2013Review, Montag2017Impact}. This approach keeps
the essential acceleration mechanism---field line shortening due to island
contraction and coalescence---and can describe the evolution of trapped and
passing electrons close to the reconnection $X$-line~\citep{Drake2013Power,
Montag2017Impact} and explained the generation of pressure anisotropy close
to the reconnection $X$-line well~\citep{Egedal2013Review}.
If one neglects the heat fluxes, the CGL closure based on the double-adiabatic
assumption predicts that the plasma heating can be expressed in terms of plasma
density and magnetic field strength~\citep{CGL1956}.~\citet{Montag2017Impact}
further showed that the plasma energization is due to the
$d\ln B/dt$ and $d\ln n/dt$, where $B$ is the magnetic strength and $n$ is
the plasma density. They included a finite compressibility
in the reduced kinetic equation and found that the finite compressibility helps
\textit{Fermi} acceleration, producing harder power-law spectrum.

Recently, by assuming the same adiabatic invariants,~\citet{Zank2014Particle} derived a
comprehensive focused transport equation that incorporates an elaborate model with
reconnection electric field, island contraction, and island coalescence for
arbitrary particle scattering levels, and a Parker-type transport equation for
the strong scattering limit, starting from a transformed Vlasov equation~\citep{Skilling1975Cosmic}.
This equation has then been used to explain the power-law distribution of energetic
particles observed in the solar wind~\citep{Zank2014Particle} and also the
anomalous cosmic-ray (ACR) energy spectrum mediated by the reconnection processes
downstream of the heliospheric termination shock~\citep{Zank2015Diffusive}.
This approach does not assume incompressible plasma and clearly distinguishes the
three acceleration mechanisms due to the mean field and plasma flow, but its
connection with particle drifts is unclear.~\citet{LeRoux2015Kinetic} derived a
more general focused transport equation, including both mean and variance of the
reconnection fields and plasma flow and starting from the standard guiding-center
kinetic equation~\citep{Kulsrud1983Handbook, LeRoux2009Time, Webb2009Drift},
and the theory now includes both incompressible and compressible energization.
This approach clearly shows the connection between the energization due to particle
drift motions (also reconnection electric field and betatron acceleration) and
energetic particle acceleration due to the background plasma flow. However,
the relative importance between the compression acceleration and other acceleration
mechanisms is undetermined.

Recent resistive MHD simulations suggest that the compression effect is important
for reconnection, especially when the plasma $\beta$ or guide field (magnetic field
component perpendicular to the reconnecting component) is
low~\citep{Birn2012Role, Provornikova2016Plasma}.~\citet{Drury2012First} treated
the acceleration of particles in reconnection similar to the diffusive shock
acceleration and showed that compression is important for driving particle
acceleration.~\citet{Zank2014Particle, LeRoux2015Kinetic, Montag2017Impact}
have pointed out that the compression effect may be important for particle
energization in reconnection regions. These appear to be in contradiction with some
previous theories that assume the reconnection layer is
incompressible~\citep[e.g.][see also the discussion
in~\citet{Pino2015Particle}]{Drake2006Electron, Drake2013Power}.

A goal of this study is to clarify the importance of compressibility in particle
energization in the magnetic reconnection layer using fully kinetic simulations that
self-consistently evolve both low-energy ``background'' plasma and high-energy
particles.

In this paper, we use moments of the Vlasov equation to derive the energization based on the
fluid motions such as fluid compression and pressure-anisotropy-related shear effect.
This approach becomes quite useful and meaningful when the system size is large enough
(i.e. much larger than the typical kinetic scales).
Using particle-in-cell (PIC) kinetic simulations, we evaluate the relative importance of
different effects and quantify the influence of the guide field and plasma $\beta$ in these
processes. We find that compressional energization dominates the acceleration
of high-energy particles when the guide field is weak, and the compression and shear
effects become comparable when the guide field is moderate (50\% of the reconnecting
component). Changing plasma $\beta$ does not significantly alter the relative
contribution of these energization terms. In Section~\ref{sec:comp}, we show how
the compression energization and shear energization terms emerge from previous
analyses based on the currents induced by guiding-center drifts.
The fully kinetic simulations and parameters are described in Section~\ref{sec:methods}.
In Section~\ref{sec:res}, we present simulation results and analyses
for electron energization. In Section~\ref{sec:con}, we discuss the conclusions
and the implications based on our simulation results.

\section{Compressional energization and shear energization}
\label{sec:comp}
Instead of starting from the drift kinetic equation for energetic
particles, we start from the Vlasov equation for the whole particle population
in the inertial frame:
\begin{equation}
  \partial_t f_s + \frac{\vect{p}}{m_s\gamma}\cdot\nabla f_s +
  q_s\left(\vect{E} + \frac{\vect{p}}{m_s\gamma}\times\vect{B}\right)\cdot
  \nabla_p f_s = 0,
\end{equation}
where $f_s$ is the phase space density, $q_s$ is the particle
charge, $m_s$ is the particle rest mass for each species (proton or electron),
$\vect{p}$ is the particle momentum, $\gamma=\sqrt{1+p^2/(m_s^2c^2)}$ is the
Lorentz factor, and $\vect{E}$ and $\vect{B}$ are electric and magnetic fields.
To study the energization of the whole particle population, we first take the moments
of this equation and obtain the conservation laws of charge, momentum, and energy,
which are
\begin{align}
  \partial_t \rho_s + \nabla\cdot\vect{j}_s & = 0, \label{equ:continuity}  \\
  \partial_t \vect{p}_s + \nabla\cdot\tensorsym{T}_s & =
  \rho_s\vect{E} + \vect{j}_s\times\vect{B}, \label{equ:mom_cons} \\
  \partial_t\mathcal{E}_s + \nabla\cdot(c^2\vect{p}_s) & =
  \vect{j}_s\cdot\vect{E}, \label{equ:ene_cons}
\end{align}
where $\rho_s=\left<q_s\right>_s$ is the charge density,
$\vect{j}_s=\left<q_s\vect{p}/m_s\gamma\right>_s$ is the current density,
$\vect{p}_s=\left<\vect{p}\right>_s$ is the momentum density,
$\tensorsym{T}_s=\left<\vect{p}\vect{p}/m_s\gamma\right>_s$ is the stress tensor,
$\mathcal{E}_s=\left<m_sc^2\gamma\right>_s$ is the particle energy density, and
$\left<A\right>_s \equiv \int d^3p Af_s$ for a general physical quantity $A$.
By assuming that the heat flux can be neglected, we truncate the fluid equation
at the second-order moments. This is consistent with the renowned CGL
closure~\citep{CGL1956}.
Equation~\ref{equ:ene_cons} shows that particles gain energy through
$\vect{j}_s\cdot\vect{E}$. Using the momentum conservation equation to evaluate
the perpendicular component of the current density $\vect{j}_{s\perp}$ w.r.t to the
local magnetic field, we found
\begin{equation}
  \vect{j}_{s\perp} = -\frac{(\nabla\cdot\tensorsym{P}_s)\times\vect{B}}{B^2} + 
  \rho_s\frac{\vect{E}\times\vect{B}}{B^2} - n_sm_s\frac{d\vect{u}_s}{dt}\times
  \frac{\vect{B}}{B^2},
\end{equation}
where we used $n_s = \rho_s/q_s$, $\vect{u}_s=\vect{p}_s/(n_sm_s)$,
$d/dt=\partial_t + \vect{v}_s\cdot\nabla$, and
$\tensorsym{T}_s=\tensorsym{P}_s + \vect{v}_s\vect{p}_s$ with
the pressure tensor $\tensorsym{P}_s$ and the species flow velocity
$\vect{v}_s=\vect{j}_s/\rho_s$. The first term on the right is due to plasma
drift caused by the pressure gradient force, the second term is due to
$\vect{E}\times\vect{B}$ drift, and the last term is due to particle inertia.
We assume that particles are well magnetized for simplicity, which leads to
\begin{equation}
  \tensorsym{P}_s = p_{s\perp}\tensorsym{I} + (p_{s\parallel} - p_{s\perp})
  \hat{\vect{b}}\hat{\vect{b}},
\end{equation}
where $p_{s\parallel}=\left<(\vect{v}_\parallel-\vect{v}_{s\parallel})
\cdot(\vect{p}_\parallel-\vect{p}_{s\parallel}/n_s)\right>$ and
$p_{s\perp}=0.5\left<(\vect{v}_\perp-\vect{v}_{s\perp})
\cdot(\vect{p}_\perp-\vect{p}_{s\perp}/n_s)\right>$ are parallel and perpendicular
pressures w.r.t the local magnetic field, $\hat{\vect{b}}=\vect{B}/B$ is the unit
vector along the local magnetic field, and $\tensorsym{I}$ is the unit dyadic.
This description is not completely accurate in regions with weak magnetic fields
and in the diffusion region because particles are not well magnetized. However,
for sufficiently large systems, the effect of the asymmetric pressure tensor has a
minor role in the energization during reconnection~\citep{Li2017Particle}.
Then, the pressure gradient effect can be broken into
\begin{align}
  \vect{j}_{s\perp} = -\frac{\nabla p_{s\perp}\times\vect{B}}{B^2} + 
  (p_{s\parallel} - p_{s\perp})\frac{\vect{B}\times(\vect{B}\cdot\nabla)\vect{B}}{B^4} +
  \rho_s\frac{\vect{E}\times\vect{B}}{B^2} - n_sm_s\frac{d\vect{u}_s}{dt}\times
  \frac{\vect{B}}{B^2}\label{equ:jperp},
\end{align}
where the first term is due to diamagnetic drift, the second term is due to
magnetic field curvature and is proportional to the pressure anisotropy.
Equation~\ref{equ:jperp} can be reorganized as
\begin{align}
  \vect{j}_{s\perp} = p_{s\parallel}\frac{\vect{B}\times(\vect{B}\cdot\nabla)\vect{B}}{B^4} +
  p_{s\perp}\frac{\vect{B}\times\nabla B}{B^3} -
  \left[\nabla\times\frac{p_{s\perp}\vect{B}}{B^2}\right]_\perp +
  \rho_s\frac{\vect{E}\times\vect{B}}{B^2} - n_sm_s\frac{d\vect{u}_s}{dt}\times
  \frac{\vect{B}}{B^2} \label{equ:jperp_drift},
\end{align}
where the first three terms are due to curvature drift, gradient drift, and
perpendicular magnetization. Note that this expression is for the whole
particle population and a similar expression could be obtained for energetic particles
from the guiding-center drift kinetic equation~\citep{Kulsrud1983Handbook,
LeRoux2015Kinetic}.

To evaluate the energy gain $\vect{j}_{s\perp}\cdot\vect{E}_\perp$,
we use $\vect{j}_{s\perp}$ from Equation 7 and $\vect{E}_\perp=-\vect{v}_E\times\vect{B}$, where
$\vect{v}_E=\vect{E}\times\vect{B}/B^2$ is the $\vect{E}\times\vect{B}$ drift.
After some algebra, we found that
\begin{align}
  \vect{j}_{s\perp}\cdot\vect{E}_\perp & =
  \nabla\cdot(p_{s\perp} \vect{v}_E) - p_s\nabla\cdot\vect{v}_E
  - (p_{s\parallel}-p_{s\perp})b_ib_j\sigma_{ij} +
  n_sm_s\frac{d\vect{u}_s}{dt}\cdot{\vect{v}_E} \label{equ:comp_shear},
\end{align}
where $p_s=(p_{s\parallel} + 2p_{s\perp})/3$ is the effective scaler pressure,
$\sigma_{ij} = 0.5 (\partial_i v_{Ej} + \partial_j v_{Ei} -
(2\nabla\cdot\vect{v}_E\delta_{ij})/3)$ is the shear tensor
for $\vect{v}_E$. The first term on the right is the flux term that does
not contribute to the energization. We define the second term as the compressional
energization, the third term as the shear energization, and the last term as the
inertial energization. Note that~\citet{LeRoux2015Kinetic} have shown
similar energization terms (Equation 13 in their paper) by using the guiding-center
drift kinetic equation, but they did not specifically point out the role of pressure
anisotropy and fluid shear.
Since current analysis employs the same two assumptions as the CGL closure, i.e.,
neglecting heat fluxes and assuming magnetized particles, the plasma energization
shown in Equation~\ref{equ:comp_shear} is consistent with other theories based on
these assumptions~\citep[][see Appendix~\ref{app:comp}]{Montag2017Impact}.
We argue that $\vect{v}_E$ is a proper choice
of perpendicular plasma flow for studying particle energization by fluid motions. 
For a macroscopic system, $\vect{v}_E$ is the leading-order drift motion
among all drift motions in the plasma perpendicular flow \citep{Hazeltine2003Plasma}.
It has been identified as the dominant perpendicular plasma flow velocity
when deriving the transport equation for studying particle acceleration~\citep{LeRoux2015Kinetic}.
Even for a relatively small-scale system as our kinetic simulations described in
the next section, $\vect{v}_E$ provides a common flow frame for both electrons and ions.

Our goal is twofold: (1) we want to test whether Equation~\ref{equ:comp_shear}
can describe the energization processes occurring in our PIC simulations of
reconnection; (2) we want to assess the relative importance of these three processes
in electron energization. One can
calculate the overall contributions to the plasma energization by compression,
shear, and inertia to evaluate the relative importance of these terms. Furthermore,
to study their energy dependence, for each particle in the simulations,
one can calculate $\vect{v}_\parallel\cdot\vect{E}_\parallel$,
$\vect{v}_\perp\cdot\vect{E}_\perp$, $-p\nabla\cdot\vect{v}_E$,
$-(p_\parallel-p_\perp)b_ib_j\sigma_{ij}$, and $m_s(d\vect{u}_s/dt)\cdot\vect{v}_E$,
where $p_\parallel=(\vect{v}_\parallel-\vect{v}_{s\parallel})
\cdot(\vect{p}_\parallel-\vect{p}_{s\parallel})$,
$p_\perp=0.5(\vect{v}_\perp-\vect{v}_{s\perp})\cdot(\vect{p}_\perp-\vect{p}_{s\perp})$,
and $p=(p_\parallel + 2p_\perp)/3$ are the contributions of each particle to the parallel
pressure, perpendicular pressure, and scalar pressure, respectively. One may then
accumulate the single-particle quantities in a series of energy bins to examine
the energy dependence of different energization effects.
For high-energy particles, this approach is consistent with that used by the
guiding-center drift kinetic equation~\citep{LeRoux2015Kinetic}, which does not
calculate each energetic particle's contribution to the pressure but the
$\vect{v}_\parallel\cdot\vect{p}_\parallel$ and $\vect{v}_\perp\cdot\vect{p}_\perp/2$
terms. The energization terms shown in Equation~\ref{equ:comp_shear} are consistent
with the double-adiabatic theories.

\section{Numerical simulations}
\label{sec:methods}
We carry out 2D kinetic simulations using the VPIC code~\citep{Bowers2008PoP},
which is a particle-in-cell code solving Maxwell's equations and the Vlasov
equation in a fully relativistic manner. The simulations start from a force-free
current sheet with $\vect{B}=B_0\tanh(z/\lambda)\hat{x} +
B_0\sqrt{\sech^2(z/\lambda) + B_g^2/B_0^2}\hat{y}$, where $B_0$ is the strength
of the reconnecting magnetic field, $B_g$ is the strength of the guide field and
$\lambda$ is the half-thickness of the current sheet.
We choose $\lambda=d_i$ in all simulations, where $d_i=c/\omega_\text{pi}=c/\sqrt{4\pi n_ie^2/m_i}$
is the ion inertial length. A reduced proton to electron mass ratio
$m_i/m_e=25$ is used for all cases. The initial particle distributions are Maxwellian
with uniform density $n_0$ and temperature $T_i=T_e=T_0$. Electrons drift
with a velocity $U_e$ that satisfies the Ampere's law.  We vary plasma
$\beta=8\pi nk(T_e+T_i)/B_0^2$ by varying $B_0$ only, which will also change the
Alfv\'en speed $v_\text{A}=B_0/\sqrt{4\pi n_0m_i}$. The electron beta
$\beta_e=8\pi nkT_e/B_0^2$ ranges from 0.02 to 0.32. The guide field strength
$B_g$ is changed from 0 to $B_0$. The parameters are listed in Table~\ref{tbl:list_runs},
which gives $c/v_\text{A}$, $c/v_\text{the}$, $\omega_\text{pe}/\Omega_\text{ce}$,
$\beta_e$ and $B_g/B_0$. We separate the runs into two groups: B1--3
indicate three runs with different plasma $\beta_e=0.02-0.32$; G1--4 indicate four runs with
$B_g=0-B_0$. The domain sizes are $L_x\times L_z=200d_i\times 100d_i$
for all simulations. The grid sizes are $4096\times2048$ for runs with $\beta_e=0.02$,
$2048\times1024$ for $\beta_e=0.08$, and $1024\times512$ for $\beta_e=0.32$.
We use 200 particles per cell per species in the runs with
$\beta_e=0.02$, 400 for $\beta_e=0.08$, and 800 for $\beta_e=0.32$.
For electric and magnetic fields, we employ periodic boundaries along the
$x$-direction and perfectly conducting boundaries along the $z$-direction.
For particles, we employ periodic boundaries along the $x$-direction and
reflecting boundaries along the $z$-direction. Initially, a long wavelength
perturbation with $B_z=0.03B_0$ is added to induce reconnection~\citep{Birn2001Geospace}.

\begin{deluxetable}{cccccccccc}
  \tabletypesize{\scriptsize}
  \tablecaption{List of simulation runs\label{tbl:list_runs}}
  \tablewidth{0pt}
  \tablehead{
  \colhead{Run} &
  \colhead{$c/v_\text{A}$} &
  \colhead{$c/v_\text{the}$} &
  \colhead{$\omega_\text{pe}/\Omega_\text{ce}$} &
  \colhead{$\beta_e$} &
  \colhead{$B_g/B_0$}
  }
  \startdata
  B1/G1 & 5.0  & 7.07 & 1.0 & 0.02 & 0.0\\
  B2    & 10.0 & 7.07 & 2.0 & 0.08 & 0.0\\
  B3    & 20.0 & 7.07 & 4.0 & 0.32 & 0.0\\
  G2    & 5.0  & 7.07 & 1.0 & 0.02 & 0.2\\
  G3    & 5.0  & 7.07 & 1.0 & 0.02 & 0.5\\
  G4    & 5.0  & 7.07 & 1.0 & 0.02 & 1.0\\
  \enddata
  \tablecomments{
    $v_\text{A}=B_0/\sqrt{4\pi n_0m_i}$ is the Alfv\'en speed of the inflow region.
    $v_{\text{the}} = \sqrt{2kT_e/m_e}$ is the electron thermal speed.
    $\omega_\text{pe}=\sqrt{4\pi n_0e^2/m_e}$ is the electron plasma frequency.
    $\Omega_\text{ce}=eB/(m_ec)$ is the electron gyrofrequency.
    $\beta_e=8\pi n_0 kT_e / B_0^2$ is the electron plasma $\beta$ based on the
    reconnection component of the magnetic field. $B_g$ is the guide field
    component of the magnetic field. B1--3 indicate runs with different
    plasma $\beta$. G1--4 indicate runs with different guide fields.
  }
\end{deluxetable}

\section{Results}
\label{sec:res}
\subsection{Compression and shear regions}
\begin{figure}[htbp]
  \centering
  \includegraphics[width=\textwidth]{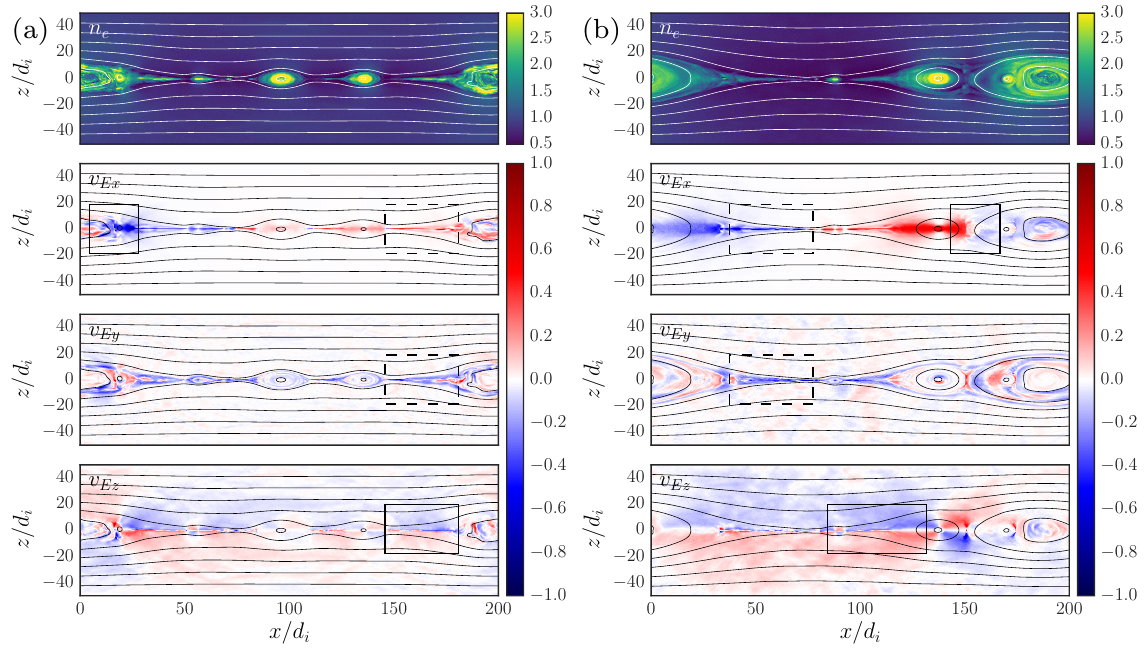}
  \caption{
    \label{fig:ne_vel}
    Electron density and three components of the $\vect{v}_E$ in
    run B1 ($\beta_e=0.02$, $B_g=0$) at $t\Omega_\text{ci}=150$ (left) and 300 (right).
    $n_e$ is normalized by the initial density $n_0$.
    The velocity components are normalized by the upstream Alfv\'en speed.
    The boxes with solid outlines indicate representative regions with fluid
    compression, and boxes with dashed outlines indicate regions with velocity shear.
  }
\end{figure}

First, we describe regions with strong compression and shear, and their
evolutions in our reconnection simulations.
Figure 1 shows the electron density $n_e$ and three components of $\vect{v}_E$
($v_{Ex}$, $v_{Ey}$, and $v_{Ez}$) at two time frames (a) $t \Omega_\text{ci} = 150$
and (b) $t \Omega_\text{ci} = 300$ for Run B1/G1. As reconnection evolves, the current
sheet breaks into a series of magnetic islands. During this process, the electron
density can increase to over three times of the initial value in reconnection exhausts
and magnetic islands. The enhanced density indicates that plasma is compressed
in the reconnection layer. The illustrated $\vect{v}_E$ components
in Figure~\ref{fig:ne_vel} further demonstrate this. We find that reconnection
exhausts, contacting islands, and island coalescence regions are the most important
regions with strong compression.
For example, the $v_{Ex}$ panels
show that the reconnection outflow is compressed in the island-merging regions;
the $v_{Ez}$ panels show that the reconnection inflow forms a compressed region at the
center of the reconnection exhaust (both are indicated by boxes with solid outlines),
leading to an enhanced electron density. Besides being compressed, the bulk flow is
also experiencing strong shear at the reconnection exhaust boundaries
and centers due to the gradient of $v_{Ex}$ and $v_{Ey}$ along the $z$-direction (boxes with
dashed outlines). As we will show below, these compressed and sheared flows can lead
to significant particle energization.

\subsection{Electron Energy Spectra and Bulk Energization due to Compression and Shear}
\begin{figure}[htbp]
  \centering
  \includegraphics[width=\textwidth]{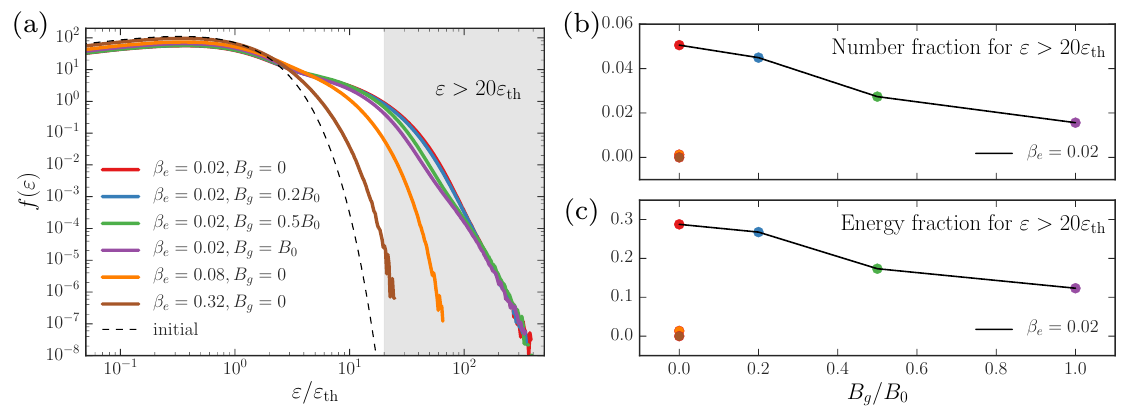}
  \caption{
    \label{fig:high_ene}
    (a) Electron energy spectra for all runs at $t\Omega_\text{ci}=600$.
    $\varepsilon_\text{th}$ is the initial thermal energy. The dashed line
    shows the initial thermal distribution, which is the same for all runs.
    The shaded region indicates the electron distribution with
    $\varepsilon > 20\varepsilon_\text{th}$.
    (b) Number fraction of the electrons with $\varepsilon > 20\varepsilon_\text{th}$
    among all electrons in the simulation box. The symbols are color-coded the same
    as those in (a). The solid line indicates the runs with $\beta_e=0.02$.
    The orange and brown symbols at the bottom left corner indicate runs with
    $\beta_e=0.08$ and 0.32, respectively.
    (c) Energy fraction of the electrons with $\varepsilon > 20\varepsilon_\text{th}$.
  }
\end{figure}

Next, we consider the details of electron energization.
Electrons are accelerated to higher energies during the reconnection processes.
Figure~\ref{fig:high_ene} (a) shows the electron energy spectra at
$t\Omega_\text{ci}=600$ in all of our simulations. High-energy tails
(kinetic energy $\varepsilon>$ 20 times that of the initial thermal energy
$\varepsilon_\text{th}$) develop in all runs, and they are more
prominent in low-$\beta$ runs than runs with higher $\beta$. The high-energy tail extends to
up to $70\varepsilon_\text{th}$ for the run with $\beta_e=0.08$ and only
$25\varepsilon_\text{th}$ for the run with $\beta_e=0.32$, and both the particle
number and particle kinetic energy in the tails in these two runs are much less
than 1\% of those quantities in all of simulations
(Figure~\ref{fig:high_ene} (b) and (c)). In contrast, for
runs with $\beta_e=0.02$, the high-energy tails extend up to $400\varepsilon_\text{th}$,
contain 1.6--5.0\% of electrons by number (Figure~\ref{fig:high_ene} (b)), and
account for 12--29\% of the energy in all of electron distributions
(Figure~\ref{fig:high_ene} (c)) depending on the guide field strength.
Figure~\ref{fig:high_ene} (a) shows that the high-energy particle flux decreases
with the guide field strength in $20\varepsilon_\text{th}<\varepsilon<100\varepsilon_\text{th}$
and that the fluxes are almost the same for electrons with
$\varepsilon>100\varepsilon_\text{th}$. Our results show that electrons with
$\varepsilon>100\varepsilon_\text{th}$ only account for less than 0.004\% of all
electrons, so we focus on the energy range $20\varepsilon_\text{th}<\varepsilon<100\varepsilon_\text{th}$,
which is statistically more important in the following discussions.

\begin{figure}[htbp]
  \centering
  \includegraphics[width=0.6\textwidth]{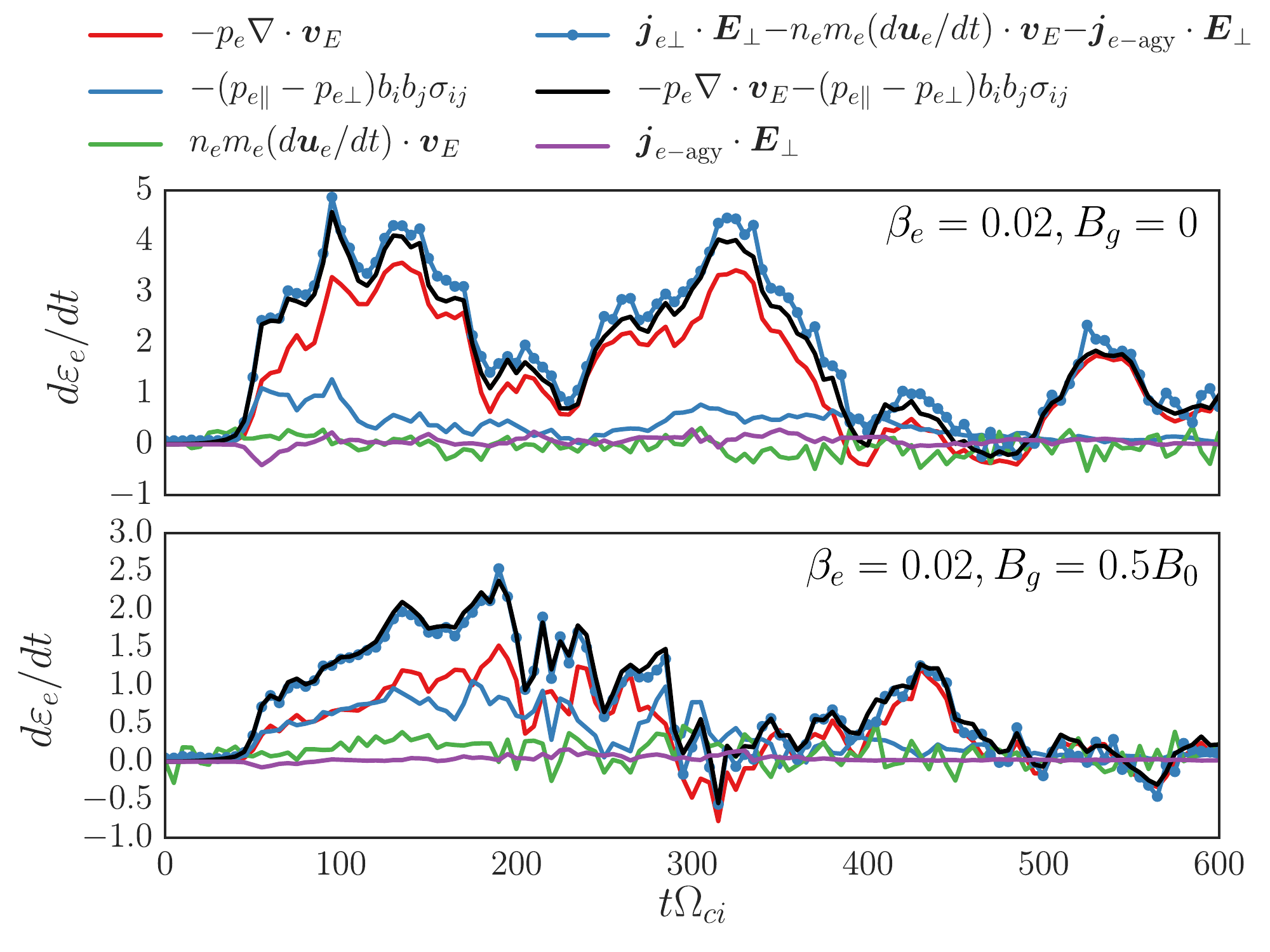}
  \caption{
    \label{fig:ene_terms}
    Time evolution of electron energization
    terms: compressional energization $-p_e\nabla\cdot\vect{v}_E$,
    shear energization $-(p_{e\parallel} - p_{e\perp})b_ib_j\sigma_{ij}$, inertial
    energization $n_em_e(d\vect{u}_e/dt)\cdot\vect{v}_E$, and agyrotropic
    energization $\vect{j}_{e-\text{agy}}\cdot\vect{v}_E$. The summation of 
    $-p_e\nabla\cdot\vect{v}_E$ and $-(p_{e\parallel} - p_{e\perp})b_ib_j\sigma_{ij}$
    (black) is compared with the energization due to perpendicular electric field
    $\vect{j}_{e\perp}\cdot\vect{E}_\perp$ subtracting the inertial
    and agyrotropic terms (blue with dots). Top panel: simulation without a guide
    field. Bottom panel: similar to the top panel, but for the simulation with a guide 
    field $\sim$ 50\% of the reconnecting component.
  }
\end{figure}

\begin{figure}[htbp]
  \centering
  \includegraphics[width=\textwidth]{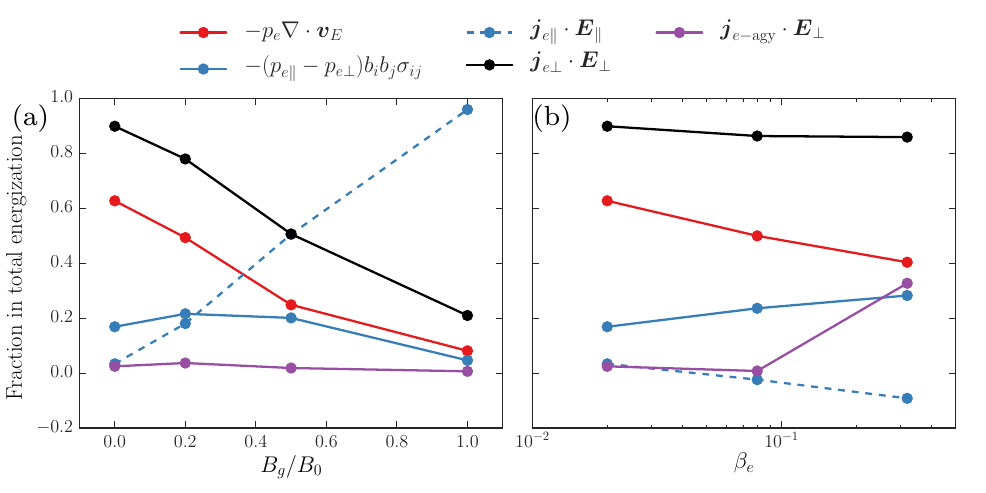}
  \caption{
    \label{fig:comp_beta_bg}
    Energization terms for runs with different guide field $B_g$ (a) and
    plasma $\beta$ (b). The energization terms are integrated over the whole simulation
    box and time until $t_1\Omega_\text{ci}=600$ and then normalized by the total particle
    energy gain at $t_1$. For example, $\vect{j}_{e\parallel}\cdot\vect{E}_\parallel$
    represents $\int d^3r \int^{t_1}_0 dt(\vect{j}_{e\parallel}\cdot\vect{E}_\parallel)/
    \Delta K_e(t_1)$. Note that the contributions tend to be underestimated
    due to the accumulated integration errors over time.
    See Table~\ref{tbl:list_runs} for the parameters of those runs.
  }
\end{figure}

To quantify the energization due to compression and shear effects, we calculate
energization due to different fluid-motion terms such as compression, shear,
and fluid inertia discussed in Section~\ref{sec:comp} (Equation~\ref{equ:comp_shear}),
as well as the contribution from agyrotropic particle distribution~\citep{Li2017Particle}.
In Figure~\ref{fig:ene_terms} we show the time evolution of each energization effect
for runs G1 ($\beta_e=0.02$, $B_g=0$) and G3 ($\beta_e=0.02$, $B_g=0.5B_0$).
The summation of different energization mechanisms (black line) agrees well with
the energization due to the perpendicular electric field $\vect{j}_\perp \cdot
\vect{E}_\perp$ (blue line with dots). The compressional
energization (red line) is dominant when there is no guide field but becomes
comparable to the shear energization (blue line) when $B_g=0.5B_0$. The inertia
term is negligible for electrons but becomes important for ions. (We will report
the energization of ions elsewhere.) As the guide field gets stronger, both
compression and shear terms are
suppressed, and the energization due to parallel electric field dominates 
(Figure~\ref{fig:comp_beta_bg} (a)). We found that the partition of these energization
terms is similar in simulations with higher plasma $\beta$
(Figure~\ref{fig:comp_beta_bg} (b)), though the compression
energization contributes less when $\beta$ is high because plasma is less 
compressible~\citep{Birn2012Role}. Another noticeable difference
is that the energization in run B3, which has the highest $\beta_e=0.32$, has
a large contribution from the nongyrotropic effects, suggesting that electrons
are not well magnetized when plasma $\beta$ is high.

\subsection{Spatial distribution of compression energization and shear energization}
\begin{figure}[htbp]
  \centering
  \includegraphics[width=\textwidth]{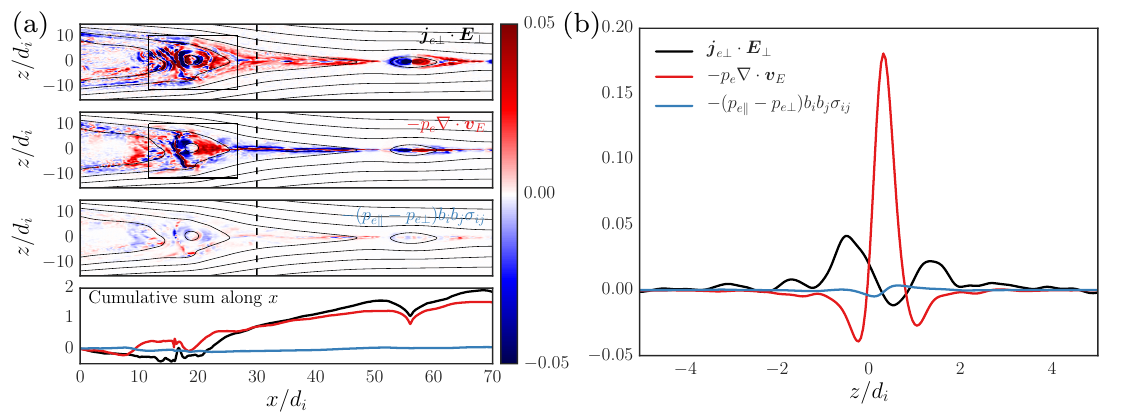}
  \caption{
    \label{fig:comp_2d_g1}
    (a) The energization terms in run G1 ($\beta_e=0.02$, $B_g=0$) at
    $t\Omega_\text{ci}=150$. The top three panels are energization due to perpendicular
    electric field, compression, and shear, respectively. The bottom panel shows
    the cumulative sum of these terms along the $x$-direction.
    The dashed lines indicate a cut along the $z$-direction.
    The energization terms are normalized by $en_0v_\text{A}^2B_0$, where $n_0$ is the
    initial electron number density, $v_\text{A}$ is the Alfv\'en speed of the inflow
    plasma, and $B_0$ is the asymptotic magnetic field strength.
    In the boxed region, a smaller island on the right is merging with the large
    island. (b) The profile of the energization terms along the dashed lines in the
    left panels. The difference between perpendicular energization and the sum
    of compression and shear energizations is due to a flux term.
    See the text for a discussion.
  }
\end{figure}

Spatial distributions of different energization effects reveal that reconnection exhausts, contracting islands,
and island-merging regions are the three most important regions for compression and
shear acceleration. Figure~\ref{fig:comp_2d_g1} shows the energization terms
in these regions in run G1 ($\beta_e=0.02$, $B_g=0$) at $t\Omega_\text{ci}=150$.
The bottom panel of Figure~\ref{fig:comp_2d_g1} (a) shows that compressional
energization (red) is the
dominant term in these regions. In the contracting island ($x\sim 57d_i$),
the compressional energization dominates as the energization primarily comes
from the converging $v_{Ex}$ and $v_{Ez}$. Detailed analysis shows that the
converging $v_{Ex}$ only contributes about 10\% of the energization in the 
contracting island and that most of the energization is through converging $v_{Ez}$.
We find that as the island moves leftward and interacts with the 
background plasma, $v_{Ez}$ slightly diverges at the left-hand side of the island due to
expansion along the $z$-direction, but the converging inflow $v_{Ez}$ on the right-hand
side contributes more, leading to a net energization.
In the region of two merging islands (boxed region in Figure~\ref{fig:comp_2d_g1}),
compressional energization dominates and peaks at the right-hand side of the smaller island
($x\sim20d_i$), where the reconnection outflow compresses the plasma in the island.
Besides magnetic islands, the reconnection exhaust is also efficient at energization,
and the compressional energization dominates in these regions (e.g. $x\sim30-50d_i$ in
Figure~\ref{fig:comp_2d_g1}). We find that $\vect{j}_{e\perp}\cdot\vect{E}_\perp$ spread
throughout the whole region of a reconnection exhaust, while the energization due
to compression is negative in most of the region, but is positive and peaks at
the center $z = 0$ (Figure~\ref{fig:comp_2d_g1}(b))
where $v_{Ez}$ switches directions (Figure~\ref{fig:ne_vel} (a), bottom panel).
The difference between these two terms is due to the flux term
$\nabla\cdot(p_{e\perp}\vect{v}_E)$, which gives zero energization in a closed
system as in our simulations.

Figure~\ref{fig:comp_2d_g1} also shows that compressional energization is nonuniform
and is accompanied with expansion in some regions.
In the anti-reconnection layer ($x\sim 18d_i$), where these two islands merge,
the overall compressional energization is small compared with other regions due
to two reasons: the convergence of $v_{Ex}$ is accompanied by the divergence  of
the outflow in the anti-reconnection region along the $z$-direction; the compression in 
the island on the right is accompanied by the expansion in the one on the left.
More detailed trajectory analyses \citep[e.g.][]{Li2017Particle} have found that some 
particles can get efficiently accelerated by accessing those compression regions.

\begin{figure}[htbp]
  \centering
  \includegraphics[width=0.5\textwidth]{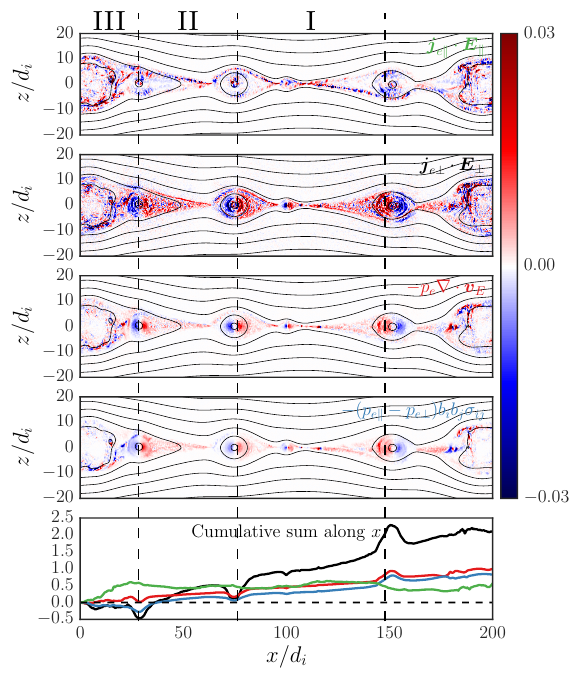}
  \caption{
    \label{fig:comp_2d_g3}
    Energization terms in run G3 ($\beta_e=0.02$, $B_g=0.5B_0$) at
    $t\Omega_\text{ci}=150$. The top four panels are energization terms by the parallel
    electric field, perpendicular electric field, compression, and shear.
    The bottom panel shows the cumulative sum of these terms along the $x$-direction.
    The vertical dashed lines separate three regions. Regions I and II are main
    reconnection sites, where the energization terms due to compression and
    shear dominate. Region III is a merging region of two magnetic islands, where
    the energization due to parallel electric field dominates. In region II,
    other terms are dominant. The energization terms are normalized by $en_0v_\text{A}^2B_0$,
    where $n_0$ is the initial electron number density, $v_\text{A}$ is the Alfv\'en
    speed of the inflow plasma, and $B_0$ is the asymptotic magnetic field strength.
  }
\end{figure}

Figure~\ref{fig:comp_2d_g1} suggests that shear energization is much weaker than
compressional energization in those regions. More analyses have shown that the 
shear energization effect is weak in most regions either because the anisotropy
is weak (e.g., in exhaust centers due to phase mixing;~\citep{Egedal2015Double})
or the shear term associated with magnetic field $b_ib_j\sigma_{ij}$ is small
(e.g., along separatrix). Shear energization becomes comparable with the compressional
energization when the guide field gets stronger (Figure~\ref{fig:comp_beta_bg}(a)).
Figure~\ref{fig:comp_2d_g3} shows the energization terms in run G3 ($\beta_e=0.02$,
$B_g=0.5B_0$) at $t\Omega_\text{ci}=150$.
We find that the energization terms due to parallel electric field, compression,
and shear are comparable but they peak in different regions. Those different effects
accelerate particles in various locations. The parallel electric field accelerates
particles along one side of the separatice~\citep{Pritchett2006Relativistic}.
The compressional energization and shear energization are comparable in reconnection
exhausts and magnetic islands but largely cancel each other in the anti-reconnection sites. 
The compressional energization is suppressed when compared with that in the run without
a guide field (Figure~\ref{fig:comp_2d_g1}), while the shear term increases due to stronger
pressure anisotropy in simulations with a higher guide field~\citep{Le2013Regimes}.
The energization due to the parallel electric field is localized close to the main
reconnection sites (e.g. region I in Figure~\ref{fig:comp_2d_g3}) and the
anti-reconnection sites (e.g. region III in Figure~\ref{fig:comp_2d_g3}),
where electrons are already energetic due to compressional energization
and shear energization at earlier stages. 

\subsection{Energy dependence of compression energization and shear energization}
\begin{figure}[htbp]
  \centering
  \includegraphics[width=\textwidth]{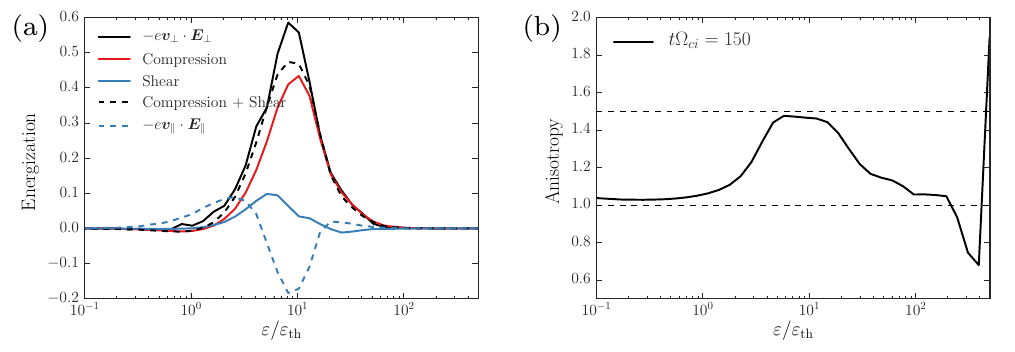}
  \caption{\label{fig:comp_ptl_g1}
    (a) The energy dependence of particle energization due to parallel electric field,
    perpendicular electric field, compression, and shear at
    $t\Omega_\text{ci}=150$ for run B1/G1 ($\beta_e=0.02$, $B_g=0$).
    The black dashed line indicates the sum of the energization due to compression
    and shear. $\varepsilon_\text{th}$ indicates the initial thermal energy.
    (b) Anisotropy for electrons with different energies.
    The anisotropy is defined as
    $\sum (\vect{v}_\parallel-\vect{v}_{s\parallel})
    \cdot(\vect{p}_\parallel-\vect{p}_{s\parallel})/\sum
    (0.5(\vect{v}_\perp-\vect{v}_{s\perp})\cdot(\vect{p}_\perp-\vect{p}_{s\perp}))$,
    where we sum over all electrons in an energy bin. The two dashed lines
    indicate anisotropy levels 1.0 and 1.5, in which 1.0 indicates the distribution
    is isotropic. Note that the peak at the highest energy bin is due to
    statistical error generated by merely a few electrons.
  }
\end{figure}

To characterize how these energization terms depend on particle energies, we
calculate the contributions of individual particles according to different
energization effects as described in Section~\ref{sec:comp} and accumulate them
in a range of energy bins to obtain the distributions of these energization
terms as a function of particle energy. Figure~\ref{fig:comp_ptl_g1} (a) and (b) show
different energization effects and anisotropy as a function of energy at
$t\Omega_\text{ci}=150$ in run B1/G1 ($\beta_e=0.02$, $B_g=0$).
Compressional energization dominates particle acceleration except for particles
at low energies ($\sim $ initial thermal energy $\varepsilon_\text{th}$). Those
low-energy particles are energized close to $X$-points by the parallel electric
field $E_\parallel$. Surprisingly, for particles with intermediate energies
($\sim 10\varepsilon_\text{th}$), the parallel electric field gives a 
cooling effect, and shear energization gives non-negligible
acceleration. For high-energy particles ($>20\varepsilon_\text{th}$),
compressional energization dominates, while the other two terms are negligible.
Shear energization is ineffective for high-energy electrons because
it requires anisotropy (Equation 9) but the anisotropy for these high-energy
electrons is less than 1.2 as shown in Figure~\ref{fig:comp_ptl_g1} (b).

\begin{figure}[htbp]
  \centering
  \includegraphics[width=\textwidth]{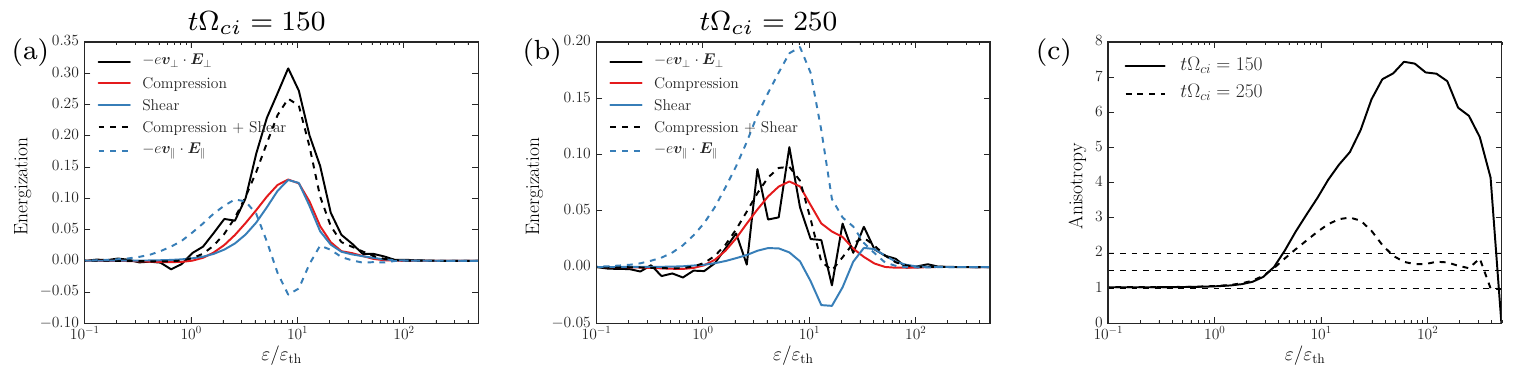}
  \caption{\label{fig:comp_ptl_g3}
    Energy dependence of particle energization due to parallel electric field,
    perpendicular electric field, compression, and shear at
    (a) $t\Omega_\text{ci}=150$ and (b) $t\Omega_\text{ci}=250$ for run G3
    ($\beta_e=0.02$, $B_g=0.5B_0$).
    The black dashed line indicates the sum of the energization due to compression
    and shear. $\varepsilon_\text{th}$ indicates the initial thermal energy.
    (c) Anisotropy for electrons with different energies.
    The anisotropy is defined as
    $\sum (\vect{v}_\parallel-\vect{v}_{s\parallel})
    \cdot(\vect{p}_\parallel-\vect{p}_{s\parallel})/\sum
    (0.5(\vect{v}_\perp-\vect{v}_{s\perp})\cdot(\vect{p}_\perp-\vect{p}_{s\perp}))$,
    where we sum over all electrons in an energy bin. The solid line is for
    $t\Omega_\text{ci}=150$ frame. The thick dashed line is for $t\Omega_\text{ci}=250$ frame.
    The three thin dashed lines indicate anisotropy levels 1.0, 1.5, and 2.0,
    in which 1.0 indicates that the distribution is isotropic.
  }
\end{figure}

The relative importance of the different energization mechanisms changes with the guide field strength.
Figure~\ref{fig:comp_ptl_g3} shows the distributions of these energization terms
at $t\Omega_\text{ci}=150$ (a) and 250 (b) for run G3 ($\beta_e=0.02$, $B_g=0.5B_0$).
At $t\Omega_\text{ci}=150$ (Figure~\ref{fig:comp_ptl_g3} (a)), compressional energization
and shear energization are comparable for particles at different energies.
This is because compressional energization is suppressed due to weak
compressibility and shear energization is enhanced due to the large anisotropy
(solid line in Figure~\ref{fig:comp_ptl_g3} (c)) when there is a finite guide field.
At the same time,
the parallel electric field accelerates low-energy electrons, decelerates intermediate-energy
electrons, and accelerates high-energy electrons, but the energization is weaker
than the other two terms. At $t\Omega_\text{ci}=250$, the parallel electric field dominates the
energization (Figure~\ref{fig:comp_ptl_g3} (b)) because it
accelerates more high-energy electrons at the island-merging regions. The relative
importance of each energization mechanism can be time variable (see below), as
the newly formed current sheet breaks into islands and major island coalescence
occurs when the islands interact with the large island as a consequence of our
periodic simulation domain. However, even though the parallel electric field dominates
the acceleration of high-energy particles at later stages in our setup, shear
energization is still larger than the energization due to $E_\parallel$ for
particles at very high energies ($>50\varepsilon_\text{th}$).
Shear energization is important for these electrons because they have
a fairly large anisotropy (> 1.5, thick dashed line in Figure~\ref{fig:comp_ptl_g3}
(c)) compared with that in the run without a guide field (Figure~\ref{fig:comp_ptl_g1} (b)).
These results show that compressional energization and shear energization are still
important for producing energetic electrons in reconnection with a moderate guide field
($B_g=0.5B_0$), while the parallel electric field becomes more important.

\subsection{Time evolution of compression energization and shear energization}
\begin{figure}[htbp]
  \centering
  \includegraphics[width=\textwidth]{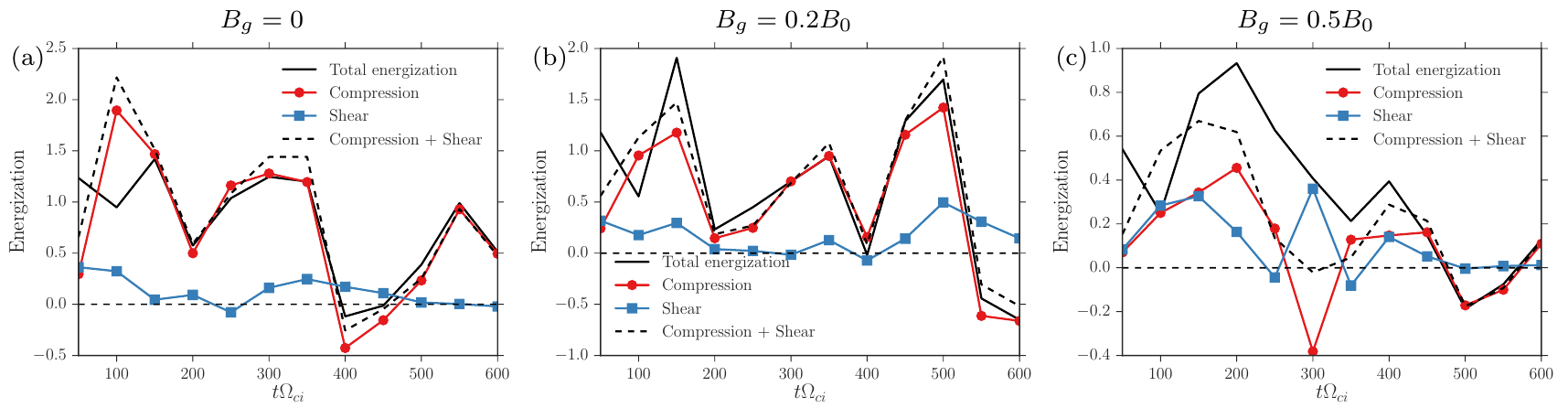}
  \caption{
    \label{fig:comp_ptl_time}
    Time evolution of compressional energization and shear energization for
    high-energy electrons ($>20$ times of the initial thermal energy) in (a) run
    B1/G1, (b) run G2, and (c) run G3. The black dashed line is the sum of these
    two terms. The black solid line is the total energization by summing
    $e\vect{v}\cdot\vect{E}$ over all high-energy electrons. Run G4 ($B_g=B_0$)
    is not shown here because the parallel energization dominates throughout
    the simulation. Run B2 ($\beta_e=0.08$) is not shown because its energization
    process is similar to run B1/G1. Run B3 ($\beta_e=0.32$) is not shown because
    very few particles can be accelerated to over 20 times of the initial thermal
    energy.
  }
\end{figure}

Figure~\ref{fig:comp_ptl_time} shows the time evolution of different
energization terms for high-energy electrons ($>20\varepsilon_\text{th}$)
in three runs with different guide fields (a) $B_g = 0$, (b) $B_g = 0.2 B_0$, and
(c) $B_g = 0.5 B_0$. In simulations without a guide field (Figure~\ref{fig:comp_ptl_time}
(a)) or with a weak guide field (Figure~\ref{fig:comp_ptl_time} (b)), compressional
energization dominates throughout the simulation. In simulations with a moderate guide field
(Figure~\ref{fig:comp_ptl_time} (c)), compressional energization and shear energization are
comparable. The sum of these
two terms contribute over 80\% of the energization at the beginning of the simulation
($t\Omega_\text{ci} < 200$), when the main reconnection layer (excluding the largest island at
the left and right boundaries) is the major energization site (Figure~\ref{fig:comp_2d_g3}).
The energization due to parallel electric field contributes over
70\% of the total energization at $200<t\Omega_\text{ci}<350$, when smaller islands
(at $x=75d_i$ and $150d_i$ in Figure \ref{fig:comp_2d_g3}) merge with the largest island.
As discussed in the previous paragraph, the time variation is likely dependent on
the detailed plasma dynamics such as the development of new sheet, island formation,
and island coalescence in a cyclic way. In a more realistic setup with open
boundaries~\citep{Daughton2006Fully}, magnetic islands grow and are then ejected out of
the system. In that situation, we expect a more evenly distributed energization over time.

\section{Discussion and Conclusion}
\label{sec:con}
In this work, we have studied the particle energization in magnetic reconnection and 
demonstrated that the energization associated with particle drift motions
can be described as energization processes due to fluid compression and shear,
especially when the system size is large enough.
The shear energization is associated with an anisotropic particle velocity distribution.
By means of fully kinetic simulations, we find that the compressional energization
dominates the energization processes in reconnection in a low-$\beta$ plasma
with a weak guide field ($\leq 0.2$ times the reconnecting component) and becomes
comparable with shear energization in reconnection with a moderate guide field
(50\% of the reconnecting component); the sum of these two terms dominates
the acceleration of high-energy particles ($>20$ times of the initial thermal energy)
except in the case with a strong guide field, in which the acceleration due to the
parallel electric field dominates.

Our analyses have shown that the compressional energization is associated with fluid
compression along both the reconnection inflow and outflow directions. We find that
the compressional energization is suppressed in simulations with an increasing guide field
and the shear energization is not suppressed until the guide field is comparable to the
reconnecting magnetic field ($B_g=B_0$). The 2D plots
(Figure~\ref{fig:comp_2d_g1} and \ref{fig:comp_2d_g3}) show that the compressional
energization and shear energization are not cospatial with the previously studied
energization term $\vect{j}\cdot\vect{E}$ because a flux term
$\nabla\cdot(p_{s\perp}\vect{v}_E)$ was not
considered~\citep{Li2015Nonthermally, Li2017Particle}. We find that
the inertial energization term is small compared with other terms for
electrons because of small electron mass but can contribute over $20\%$ of
energization for ions~\citep{Li2017Particle}. We will discuss its effect on high-energy
ion acceleration in a future study.

The connection between particle drifts and compression is consistent with previous
results in energetic particle transport theory~\citep{Jokipii1982Particle,
Jones1990Generalized,LeRoux2015Kinetic}.
Our results on energization processes are consistent with~\citet{Birn2012Role},
who performed MHD simulations and demonstrated that fluid compression is the leading
mechanism for plasma energization in low-$\beta$ plasma with a low guide field. These
results differ from some previous modeling works, in which the authors assumed
that the reconnection layer is incompressible~\citep{Bian2013Stochastic, Drake2013Power}.
Compressibility has been emphasized in recent models of particle
energization in magnetic reconnection~\citep{LeRoux2015Kinetic,Montag2017Impact}. This
work provides the first quantitative evaluation of the role of compressibility in
fully kinetic simulations.
Also, the plasma energization described by Equation~\ref{equ:comp_shear}
is consistent with the general analytical theory by
\citet[][see Appendix~\ref{app:comp}]{Montag2017Impact} based on double-adiabatic
assumptions.

The anisotropic momentum distribution of energetic particles is important for
shear acceleration. Our 2D kinetic simulations show that this leads to non-negligible
acceleration when a moderate guide field exists. The anisotropic distribution can be
generated  by electron trapping~\citep{Egedal2013Review} and curvature/gradient drift
motions~\citep{Drake2010Magnetic,LeRoux2015Kinetic}.
The anisotropy tends to be weakened when the particle orbits are chaotic in the weak
guide field limit or if strong wave-particle interaction presents. Quantifying
the role of anisotropic distribution in energetic particle acceleration in the reconnection
region is an important problem for future studies.

Our 2D kinetic simulations have a few limitations. First, we are forced to
use a relatively low mass-ratio $m_i/m_e = 25$ in order to capture the long-term
energy conversion in low-$\beta$ reconnection with a fairly large simulation
domain, but the plasma dynamics and field structures might change
with the mass ratio, especially for simulations with a guide field~\citep{Le2013Regimes}. The
second limitation is that the 2D configuration prevents the gradient of fluid
velocity along the out-of-plane direction, and this might 
influence the energization due to fluid compression and shear. Also, a real 3D configuration
leads to the development of turbulence~\citep{Bowers2007Spectral,Daughton2011Role,
Liu2011Particle, Guo2015Particle}, which can scatter particles and reduce pressure
anisotropy. Another limitation is that the drift analysis does not include compression
of fluid velocity along the magnetic field direction. This is usually achieved through
wave-particle interaction and is out of the scope of the current study. We defer
these studies to a future work.

To conclude, we find that the compressional energization and shear energization
are the major mechanisms for high-energy particle acceleration during reconnection in
a plasma with low-$\beta$ and a weak or moderate guide field and the shear
energization is proportional to the pressure anisotropy. This study links the acceleration
mechanisms found in kinetic simulations with that in energetic particle transport
theory~\citep[e.g.][]{Parker1965Passage, Drake2013Power, Zank2014Transport,
LeRoux2015Kinetic}. It provides clues for building an energetic
particle transport model for particle acceleration in solar flares and other
astrophysical reconnection sites.

\acknowledgments
We acknowledge the support by NASA under grant NNH16AC60I, DOE OFES, and the
support by the DOE through the LDRD program at LANL. F.G.'s contributions are
partly based upon work supported by the U.S. Department of Energy,
Office of Fusion Energy Science, under Award Number DE-SC0018240.  We gratefully
acknowledge our discussions with Xiangrong Fu and Andrey Beresnyak.
Simulations were performed with LANL institutional computing and also at the National
Energy Research Scientific Computing Center.

\appendix
\section{Comparison on Energization between double-adiabatic Theories and
the Current Analysis}
\label{app:comp}
By assuming that particles are magnetized and neglecting the heat
flux~\citep{CGL1956},
\citet{Montag2017Impact} showed that the energization for a single energetic
particle is [their Equation 12] 
\begin{equation}
  \frac{dU}{dt} = -mv_\parallel^2\left[\frac{\dot{B}}{B}
  \left(1-\frac{v_\perp^2}{2v_\parallel^2}\right)-\frac{\dot{n}}{n}\right].
  \label{equ:du_dt}
\end{equation}
where $U$ is the total particle energy, $m$ is the particle mass, $B$ is the
magnetic field strength, $\dot{B}\equiv dB/dt$, $n$ is the plasma density,
$\dot{n} \equiv dn/dt$, and $v_\parallel$ and $v_\perp$ are the parallel and
perpendicular particle velocities, respectively. From the continuity equation,
we get $\dot{n}/n = -\nabla\cdot\vect{V}$, where
$\vect{V}$ is plasma velocity. From the induction
equation
$\partial\vect{B}/\partial t = \nabla\times(\vect{V}\times\vect{B})$~\citep{CGL1956},
\begin{equation}
  \frac{\dot{B}}{B} = -\nabla\cdot\vect{V} + b_ib_j\frac{\partial V_i}{\partial x_j}.
\end{equation}
Integrating Equation~\ref{equ:du_dt} over the velocity space, we get the total particle
energization
\begin{equation}
  \frac{d\mathcal{E}}{dt} = -(p_\parallel - p_\perp)
  \left(-\nabla\cdot\vect{V} + b_ib_j\frac{\partial V_i}{\partial x_j}\right)
  - p_\parallel \nabla\cdot\vect{V},
\end{equation}
where $\mathcal{E}$ is the particle energy density, $p_\parallel$ and $p_\perp$
are the parallel and perpendicular pressures, respectively, and
\begin{align}
  \frac{\partial V_i}{\partial x_j} & =
  \frac{1}{3}\frac{\partial V_k}{\partial x_k}\delta_{ij} +
  \frac{1}{2}\left(\frac{\partial V_i}{\partial x_j} +
  \frac{\partial V_j}{\partial x_i} -
  \frac{2}{3}\frac{\partial V_k}{\partial x_k}\delta_{ij}\right) +
  \frac{1}{2}\left(\frac{\partial V_i}{\partial x_j} -
  \frac{\partial V_j}{\partial x_i}\right) \\
  & = \frac{1}{3}\nabla\cdot\vect{V}\delta_{ij} +
  \sigma_{ij} + \omega_{ij},
\end{align}
where $\sigma_{ij}$ is the shear tensor and $\omega_{ij}$ is the rotation tensor.
Then,
\begin{align}
  \frac{d\mathcal{E}}{dt} & = -\frac{p_\parallel + 2p_\perp}{3}\nabla\cdot\vect{V}
  - (p_\parallel - p_\perp)b_ib_j\sigma_{ij}, \\
  & = -p\nabla\cdot\vect{V} - (p_\parallel - p_\perp)b_ib_j\sigma_{ij},
\end{align}
where we used $b_ib_j\omega_{ij}=0$.
This is consistent with the dominant energization terms in
Equation~\ref{equ:comp_shear}.

\bibliography{references}{}
\bibliographystyle{aasjournal}
\end{document}